\documentclass[conference,a4paper]{IEEEtran}
\IEEEoverridecommandlockouts
\usepackage{cite}
\usepackage{amsmath,amssymb,amsfonts}
\usepackage{algorithmic}
\usepackage{graphicx}
\usepackage{textcomp}
\usepackage{makecell}
\usepackage{array}
\usepackage{booktabs}

\usepackage{amsbsy,cases,bm,subfigure}
\usepackage[ruled]{algorithm2e} 
\usepackage{xcolor}

\makeatletter
 \let\old@ps@headings\ps@headings
 \let\old@ps@IEEEtitlepagestyle\ps@IEEEtitlepagestyle
 \def\confheader#1{%
 \def\ps@headings{%
 \old@ps@headings%
 \def\@oddhead{\strut\hfill#1\hfill\strut}%
 \def\@evenhead{\strut\hfill#1\hfill\strut}%
 }%
 \def\ps@IEEEtitlepagestyle{%
 \old@ps@IEEEtitlepagestyle%
 \def\@oddhead{\strut\hfill#1\hfill\strut}%
 \def\@evenhead{\strut\hfill#1\hfill\strut}%
 }%
 \ps@headings%
 }
\makeatother

\confheader{%
This paper has been accepted for oral presentation at IPIN 2021, Lloret de Mar, Spain}

\def\BibTeX{{\rm B\kern-.05em{\sc i\kern-.025em b}\kern-.08em
    T\kern-.1667em\lower.7ex\hbox{E}\kern-.125emX}}
\begin{document}

\title{Two Efficient and Easy-to-Use NLOS Mitigation Solutions to Indoor 3-D AOA-Based Localization

\thanks{This work was supported by the state graduate funding coordinated by Uni Freiburg.\\
$^\S$\textit{Corresponding author: Wenxin Xiong} (xiongw@informatik.uni-freiburg.de).}
}

\author{\IEEEauthorblockN{Wenxin~Xiong$^{\dagger\S}$,~Joan~Bordoy$^\dagger$,~Andrea~Gabbrielli$^\ddagger$,~Georg~Fischer$^*$,~Dominik~Jan~Schott$^\ddagger$,~Fabian~H\"oflinger$^{*\ddagger}$,\\Johannes~Wendeberg$^\dagger$,~Christian~Schindelhauer$^\dagger$,~and~Stefan~Johann~Rupitsch$^\ddagger$}
\IEEEauthorblockA{$^\dagger$\textit{Department of Computer Science,~University of Freiburg, Freiburg 79110, Germany} \\
$^\ddagger$\textit{Department of Microsystems Engineering,~University of Freiburg, Freiburg 79110, Germany} \\
$^*$\textit{Fraunhofer Institute for Highspeed Dynamics, Ernst-Mach-Institute (EMI), Efringen-Kirchen 79588, Germany}}
}

\maketitle

\begin{abstract}
	This paper proposes two efficient and easy-to-use error mitigation solutions to the problem of three-dimensional (3-D) angle-of-arrival (AOA) source localization in the mixed line-of-sight (LOS) and non-line-of-sight (NLOS) indoor environments. A weighted linear least squares estimator is derived first for the LOS AOA components in terms of the direction vectors of arrival, albeit in a sub-optimal manner. Next, data selection exploiting the sum of squared residuals is carried out to discard the error-prone NLOS connections. In so doing, the first approach is constituted and more accurate closed-form location estimates can be obtained. The second method applies a simulated annealing stochastic framework to realize the robust $\ell_1$-minimization criterion, which therefore falls into the methodology of statistical robustification. Computer simulations and ultrasonic onsite experiments are conducted to evaluate the performance of the two proposed methods, demonstrating their outstanding positioning results in the respective scenarios.
	\end{abstract}

	
	\section{Introduction}
Source localization using the angle-of-arrival (AOA) measurements collected by multiple spatially separated sensors without time synchronization, especially that in the general three-dimensional (3-D) setting, has recently witnessed a research upsurge as the mobile communication, radar, sonar, wireless sensor network, and acoustic indoor localization technologies evolve \cite{AGabbrielli,FZafari,STomic_LE,NHNguyen,STomic5,STomic6}.

Similar to their distance-based counterparts \cite{Guvenc1}, angle-based localization schemes taking advantage of the source direction information (particularly, the AOA) relative to the sensors can also be badly affected by the occurrence of outliers with abnormally large values, for reasons like the unavoidable non-line-of-sight (NLOS) propagation conditions in indoor environments. This is mainly because conventional methods for AOA-based source localization often rely on the von Mises or simply Gaussian noise assumption, in view of their theoretical/computational convenience and good approximation of reality \cite{STomic_LE,NHNguyen,STomic5,STomic6}. Apparently, such algorithms cannot work reliably under the corresponding adverse circumstances.

Most studies in this field were to discuss the mitigation of NLOS errors in two-dimensional (2-D) AOA-based localization that acquires merely the one-dimensional bearing measurements as AOA observations \cite{QYan,QYan2,PGF}. Depending on how the error-prone data are treated, these methods can be roughly divided into the robust statistics/outlier detection \cite{QYan,QYan2} and expectation maximization \cite{PGF} ones. In the general 3-D setting of the localization system, mitigating the bias errors in both azimuth and elevation angle measurements has been considered by the authors of \cite{NHNguyen} and \cite{KYu}. In \cite{NHNguyen}, the $\ell_p$-norm minimization criterion is applied to 3-D AOA-based source localization in the presence of $\alpha$-stable impulsive noise, and an iterative reweighted instrumental-variable estimator (IRIVE) is designed in order to achieve the theoretical covariance. In spite of the estimation unbiasedness and performance advantages guaranteed, the authors of \cite{NHNguyen} premise their study on perfect prior knowledge of the impulsive noise distribution parameters. This might hinder the practical application of the IRIVE scheme. On the contrary, a statistical hypothesis testing approach has been developed in an earlier work \cite{KYu} to identify the NLOS links among nodes in the array networks, which nevertheless deviates from the topic of AOA-based single-source localization here.

The brief discussion above implies that there is in general a lack of adequate algorithmic solutions to 3-D AOA-based source localization using possibly unreliable sensor-collected measurements. In this paper, we continue to investigate such a problem, and focus specifically on the instances in the mixed line-of-sight (LOS) and NLOS indoor environments. As also the main contribution of the article, we take an initial step in this direction by proposing two efficient and easy-to-use NLOS error mitigation methods for 3-D AOA localization.

We follow the Gaussian-uniform mixture error modeling strategy in the 2-D study \cite{QYan}, and assume that the mixed distributions simply degenerate to Gaussian processes in the LOS scenarios. With the aim to devise a practically applicable algorithm, we derive first a weighted linear least squares (LS) method for the LOS source-sensor connections through the conversion into spherical coordinates, to which the analogues are fairly common in the literature \cite{STomic_LE,STomic6}. Despite its statistical sub-optimality, the weighted linear LS solution is obtained in the closed form and can thus be computationally very efficient. Different from the plain linear estimators in \cite{STomic_LE,STomic6}, we introduce the procedure of data selection into the framework by exploiting an LS cost function of the residuals. In doing so, only a subset of AOA measurements minimizing the loss will be picked out and utilized for the ascertainment of source location. In fact, the countermeasures of data selection have been successfully taken in the areas of direction-of-arrival estimation \cite{AMCRBor} and time-of-arrival-, time-difference-of-arrival-, and time-sum-of-arrival-based positioning \cite{PCChen,JAApo,WXiong5}, but not yet been made use of for improving the resistance of 3-D AOA-based localization algorithms to the NLOS errors.

Our second NLOS mitigation approach employs a modified simulated annealing (SA) method \cite{LIngber} to stochastically solve the nonconvex and nonsmooth $\ell_1$-minimization problem, which is justified by the widespread use, strong outlier-resistance, and low prior knowledge requirement of the least $\ell_1$ norm estimation criterion in robust source localization \cite{WXiong3,WXiong4}. Furthermore, unlike the deterministic algorithms imposing prerequisites to continuity and/or differentiability of the objective function, stochastic search methods based on the random variables do not rely on such assumptions, and can benefit from their capacity for escaping from the local optima \cite{LIngber}. These aspects have made the stochastic methodology an appealing candidate for engineering optimization tasks with intricate and multimodal cost functions, e.g., array self-calibration \cite{OJean}.

The remainder of the paper is structured as follows. Section \ref{PF} states the 3-D AOA-based localization problem to be solved and the system model. Section \ref{AD} develops the weighted linear LS estimator and describes the procedure of data selection. Section \ref{L1} derives the modified SA algorithm for $\ell_1$-minimization. Computer simulations and real-world experiments are performed in Section \ref{NR} to evaluate the performance of the proposed algorithms. Finally, Section \ref{CC} concludes the paper.

\section{Problem Formulation}
\label{PF}
\begin{figure}[!t]
	\centering
	\includegraphics[width=3.5in]{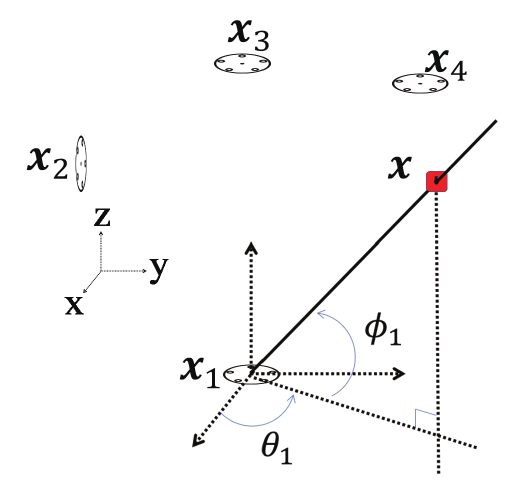}
	\caption{Geometry of 3-D AOA-based single-source localization.} 
	\label{AOAillustration}
\end{figure}
We consider here the problem of AOA-based single-source localization in the 3-D space using $L \geq 2$ spatially separated sensors equipped with antenna array or directional antenna. As depicted in Fig. \ref{AOAillustration}, our goal is to locate the source, whose unknown location is denoted by $\bm{x} = \left[ x,y,z \right]^T \in \mathbb{R}^{3}$, with the use of the known sensor positions $\bm{x}_i = \left[ x_i,y_i,z_i \right]^T \in \mathbb{R}^{3}$ (for $i = 1,...,L$) and a set of sensor-collected AOA pairs modeled as
\begin{align}{\label{NoisyAOA}}
&\hat{\theta}_i = \theta_i + m_i,~~i = 1,...,L,\nonumber\\
&\hat{\phi}_i = \phi_i + n_i,~~i = 1,...,L,
\end{align}
where $\theta_i = \textup{atan2} \left( y - y_i, x - x_i \right) \in [ - \pi, \pi]$ and $\phi_i = \textup{atan2} \left( z - z_i, {\left\| \bm{E} \left( \bm{x} - \bm{x}_i \right) \right\|}_2 \right) \in [ - \pi, \pi]$ are the true azimuth and elevation angles, respectively, $\textup{atan2}\left( Y, X \right)$ returns the four-quadrant inverse tangent of $Y$ and $X$, $\hat{\theta}_i \in [ - \pi, \pi]$ and $\hat{\phi}_i \in [ - \pi, \pi]$ are their noisy counterparts, ${\left\| \cdot \right\|}_2$ stands for the $\ell_2$-norm, $\bm{E} = [1,0,0;0,1,0]\in \mathbb{R}^{2 \times 3}$, and $m_i$ and $n_i$ are the error components in the corresponding angle measurements assumed to be independent and identically distributed.

We follow \cite{Guvenc1,QYan,WXiong,WXiong3,WXiong4,WXiongMCC} to model $m_i$ and $n_i$ in a broadly applicable way as the mixture of Gaussian and uniform distributions:
\begin{align}{\label{CF_general}}
&f \left( m_i \right) = \left( 1-p_{i} \right) \mathcal{N} \left( 0, \sigma_{i}^2 \right) + p_{i} \mathcal{U} \left( -\pi, \pi \right),\nonumber\\
&f \left( n_i \right) = \left( 1-p_{i} \right) \mathcal{N} \left( 0, \sigma_{i}^2 \right) + p_{i} \mathcal{U} \left( -\pi, \pi \right),
\end{align}
where $f(\cdot)$ represents the probability density function, $p_{i}$ denotes the probability that the $i$th sensor is in the NLOS environment, and $\mathcal{N} \left( 0, \sigma_i^2 \right)$ and $\mathcal{U} \left( -\pi, \pi \right)$ are the zero-mean Gaussian distribution with variance $\sigma_{i}^2$ and the uniform distribution within interval $[ - \pi, \pi]$, respectively. We focus herein on the more general localization scenarios without any prior information about $p_{i}$ or $\sigma_i$, and the only assumption made about the parameter settings is that $m_i$ and $n_i$ either reduce to Gaussian processes or are contaminated by some degree of NLOS errors, depending on whether the $i$th source-sensor path corresponds to LOS or NLOS, respectively. This is consistent with the existing work on robust NLOS error mitigation in the field of distance-based localization \cite{WXiong,WXiong3,WXiong4,WXiongMCC}.

\section{Data-Selective LS Solution}
\label{AD}
In this section, our residual-based data-selective LS algorithm is devised.

\subsection{Weighted Linear LS Estimator}
An easily realizable weighted linear LS estimator is derived first for the LOS source-sensor connections. Assuming that the Gaussian noise in the azimuth and elevation observations under LOS propagation is small compared with the angle measurements themselves, (\ref{NoisyAOA}) straightforwardly leads to:
\begin{subequations}
	\begin{align}
		&\hat{\bm{c}}_{i}^T \left( \bm{x} - \bm{x}_i \right) \approx 0,~~i = 1,...,L,{\label{NoisyAOA2a}}\\
		&\bm{k}^T \left( \bm{x} - \bm{x}_i \right) - {\left\| \bm{x} - \bm{x}_i \right\|} \sin \left( \hat{\phi}_i \right) \approx 0,~~i = 1,...,L,{\label{NoisyAOA2b}}
		\end{align}
\end{subequations}
where $\hat{\bm{c}}_{i} = \left[ -\sin \left( \hat{\theta}_i \right), \cos \left( \hat{\theta}_i \right), 0 \right]^T \in \mathbb{R}^3,~~i = 1,...,L,$ and $\bm{k} = \left[ 0,0,1 \right]^T \in \mathbb{R}^3$. Since the direction of source with respect to the $i$th sensor can be indicated by a unit vector (UV) known as the direction vector of arrival (DVOA) \cite{TKLe}: 
\begin{equation}{\label{DVOA_def}}
\bm{u}_i = \left[ \cos \phi_i \cos \theta_i, \cos \phi_i \sin \theta_i, \sin \phi_i \right]^T \in \mathbb{R}^3,
\end{equation}
it is able to re-express the source-sensor distance constraints $d_{i} = {\left\| \bm{x} - \bm{x}_i \right\|}_2$ (for $i = 1,...,L$) in the spherical coordinate system as
\begin{equation}{\label{SCY}}
	\bm{x} - \bm{x}_i = \bm{u}_i d_{i},~d_{i} \geq 0,~{\left\| \bm{u}_i \right\|}_2=1,~~i=1,...,L.
\end{equation}
Through the use of $\bm{u}_i$ with $\bm{u}_i^T \bm{u}_i = 1$ (for $i=1,...,L$), (\ref{NoisyAOA2b}) is equivalently written as
\begin{equation}
	\bm{k}^T \bm{u}_i d_i - \bm{u}_i^T \bm{u}_i d_i \sin \left( \hat{\phi}_i \right) \approx 0,~~i = 1,...,L,
\end{equation}
which further implies:
\begin{equation}{\label{elev_new}}
	\left( \bm{k} - \bm{u}_i \sin \left( \hat{\phi}_i \right) \right)^T \left( \bm{x} - \bm{x}_i \right) \approx 0,~~i = 1,...,L.
\end{equation}

Replacing $\bm{u}_i$ in (\ref{elev_new}) with the available UV $\hat{\bm{u}}_i = \left[ \cos \hat{\phi}_i \cos \hat{\theta}_i, \cos \hat{\phi}_i \sin \hat{\theta}_i, \sin \hat{\phi}_i \right]^T \in \mathbb{R}^3$, the source location estimate $\hat{\bm{x}}$ can be obtained by minimizing a weighted linear LS cost function based on (\ref{NoisyAOA2a}) and (\ref{elev_new}) as follows:
\begin{align}{\label{WLLS}}
	&\tilde{\bm{x}} = \arg \min_{\bm{x}} \sum_{i=1}^{L} w_i \left( \hat{\bm{c}}_{i}^T \left( \bm{x} - \bm{x}_i \right) \right)^2 \nonumber\\
	&~~+ \sum_{i=1}^{L} w_i \left( \left( \bm{k} - \hat{\bm{u}}_i \sin \left( \hat{\phi}_i \right) \right)^T \left( \bm{x} - \bm{x}_i \right) \right)^2,
\end{align}
where $w_i = 1 - \tfrac{{\left\| \bar{\bm{x}} - \bm{x}_i \right\|}_2}{\sum_{i=1}^{L} {\left\| \bar{\bm{x}} - \bm{x}_i \right\|}_2}$ (for $i=1,...,L$) are weights assigned to express stronger preference for the neighboring connections and $\bar{\bm{x}}$ denotes the initial LS estimate from (\ref{WLLS}) by setting all weights to 1. The justification for introducing $\{ w_i \}$ is that a longer source-sensor distance (radius for sensor-centered circle) implies a larger estimation deviation from the true position (chord length) for certain $m_i$ or $n_i$ (central angle) \cite{STomic_LE,STomic6}.

The weighted LS formulation in (\ref{WLLS}) can be rewritten into an equivalent vector form as
\begin{align}{\label{WLLS_vec}}
	\min_{\bm{x}} \left( \bm{A} \bm{x} - \bm{b} \right)^T \bm{W} \left( \bm{A} \bm{x} - \bm{b} \right),
\end{align}
where $\bm{W} = \bm{I}_{2} \otimes \textup{diag}(\bm{w}) \in \mathbb{R}^{2L \times 2L}$, $\bm{I}_2 \in \mathbb{R}^{2 \times 2}$ represents the identity matrix of size 2, $\bm{w} = \left[ w_1,...,w_L \right]^T \in \mathbb{R}^{L}$, $\bm{A} = \left[ \hat{\bm{c}}_{1}^T;...;\hat{\bm{c}}_{L}^T; \left( \bm{k} - \hat{\bm{u}}_1 \sin \left( \hat{\phi}_1 \right) \right)^T;...; \left( \bm{k} - \hat{\bm{u}}_L \sin \left( \hat{\phi}_L \right) \right)^T \right]$ $\in \mathbb{R}^{2L \times 3}$, and $\bm{b} = \bigg[ \hat{\bm{c}}_{1}^T \bm{x}_1;...;\hat{\bm{c}}_{L}^T \bm{x}_L; \left( \bm{k} - \hat{\bm{u}}_1 \sin \left( \hat{\phi}_1 \right) \right)^T \bm{x}_1;$ $...; \left( \bm{k} - \hat{\bm{u}}_L \sin \left( \hat{\phi}_L \right) \right)^T \bm{x}_L \bigg] \in \mathbb{R}^{2L}$. The closed-form solution to (\ref{WLLS_vec}) is simply
\begin{equation}{\label{WLLS_cfsolu}}
	\tilde{\bm{x}} = \left( \bm{A}^T \bm{W} \bm{A} \right)^{-1} \left( \bm{A}^T \bm{W} \bm{b} \right).
\end{equation}
Note that the 3-D AOA-based weighted linear LS estimator described above is not a new result, but it is normally built into localization methods considering various forms of location-bearing information in the recent literature, e.g., integrated received signal strength and AOA measurements \cite{STomic_LE,STomic6}.

\subsection{Procedure of Data Selection}
We now present our residual-based data-selective approach for circumventing the susceptibility of (\ref{WLLS_cfsolu}) to NLOS propagation. The key idea is to exploit an LS cost function of the residuals using $N$ DVOAs from the ensemble $L$ observations, defined as
\begin{align}{\label{Residual_LS}}
    &R\left( N, \tilde{x}_N \right) = \frac{1}{N} \sum_{ i \in \mathbb{S}_N} \left( \hat{\bm{c}}_{i}^T \left( \tilde{\bm{x}}_{\{ N \}} - \bm{x}_i \right) \right)^2 \nonumber\\
	&~~+ \frac{1}{N} \sum_{ i \in \mathbb{S}_N} \left( \left( \bm{k} - \hat{\bm{u}}_i \sin \left( \hat{\phi}_i \right) \right)^T \left( \tilde{\bm{x}}_{\{ N \}} - \bm{x}_i \right) \right)^2,
\end{align}
where $\mathbb{S}_{N}$ denotes a subset of $N$ DVOAs which belongs to the $N$-combination in test and is of cardinality $\textup{card}(\mathbb{S}_{N}) = N$, and $\tilde{\bm{x}}_{\{ N \}}$ represents the corresponding weighted linear LS location estimate.

Our data-selective algorithm adopts the classical identifying and discarding (IAD) strategy to rule out those error-prone combinations. The weighted linear LS solution provided in the last subsection will be tailored to combinatorially test every possibility of $\mathbb{S}_{N}$, whereafter the one producing the minimum $R\left( N, \tilde{x}_N \right)$ will be treated as the LOS set and $\tilde{x}_N$ the final location estimate. For brevity, we summarize the whole procedure in Algorithm 1.

\begin{algorithm}[t]
	\SetAlgoNoLine
	\caption{IAD Residual-Based Data Selection for NLOS Error Mitigation in 3-D AOA Localization.}
	\KwIn{Available DVOAs $\{ \hat{\bm{u}}_i \}$, sensor positions $\{\bm{x}_i\}$, and predefined $N \in \left\{ N \in \mathbb{Z} | 2 \leq N \leq L \right\}$.}
	
	{
		
		$~~$\textbf{Initialize:} $\tilde{\bm{x}}_{N}$ with linear LS estimate from (\ref{WLLS_cfsolu})
		$~~$using all $L$ DVOAs (viz., $\mathbb{S}_{L}$), and $\delta_{N} = R(L,\tilde{\bm{x}}_{N})$.
		
		$~~$\textbf{for} $i=1,2,...,\frac{L!}{N! (L-N)!}$ (viz., each $\mathbb{S}_{N}$) \textbf{do}
		
			$~~~~$Pass $N$ DVOAs in the $i$th test to the weighted 
			
			$~~~~$linear LS method to yield a position estimate $\tilde{\bm{x}}_{N}^{\{i\}}$. 
			
			$~~~~$The original definitions of $\bm{A}$, $\bm{b}$, and $\bm{W}$ in (\ref{WLLS_cfsolu}) are 
			
			$~~~~$accordingly modified. Only $2N$ rows of $\bm{A}$ and $\bm{b}$
			
			$~~~~$associated with the considered $N$ DVOAs are kept,
			
			$~~~~$whereas the rows and columns of $\bm{W}$ not associated
			
			$~~~~$with the corresponding $N$ DVOAs are removed.

			$~~~~$\textbf{if} $R\left( N, \tilde{\bm{x}}_{N}^{\{i\}} \right) < \delta_{N}$ \textbf{then} 
			
			$~~~~~~$$\delta_{N} \leftarrow R\left( N, \tilde{\bm{x}}_{N}^{\{i\}} \right)$ and $\tilde{\bm{x}}_{N} \leftarrow \tilde{\bm{x}}_{N}^{\{i\}}$.

			$~~~~$\textbf{end~if}

	    $~~$\textbf{end~for} with $\tilde{\bm{x}} = \tilde{\bm{x}}_{N}$.
		
	}
	\KwOut{Final estimate of source location $\tilde{\bm{x}}$.}
\end{algorithm}

We see that Algorithm 1 has its limitations because the number of DVOAs used for localization must be specified prior to the invocation. Comparatively speaking, the negative impacts of NLOS propagation on the positioning performance might not be thoroughly reduced if the value of $N$ is too large, whereas $N \ll N_{\textup{LOS}}$ might result in information loss to a certain extent. This implies a fine balance between the mitigation of NLOS errors and exploitation of available measurements needs to be struck. Nonetheless, mature hyperparameter tuning methods such as cross-validation \cite{WJZeng} can actually be utilized in the practical localization applications.

\section{Stochastic $\ell_1$-Minimization}
\label{L1}

Our second approach does not hinge on the choice of $N$. Here, we robustify the non-outlier-resistant LS criterion in the $\ell_2$-space by the $\ell_1$-norm-based counterpart, and propose to handle:
\begin{align}{\label{l1_robust}}
    \min_{\bm{x}} {\psi} \left( \bm{x} \right) := \sum_{i = 1}^{L} \left( \left| \hat{\theta}_i - \theta_i \right| +  \left| \hat{\phi}_i - \phi_i \right| \right).
\end{align}
Since (\ref{l1_robust}) is both nonconvex and nonsmooth, tackling it by the traditional deterministic optimization algorithms might easily get trapped into local optimum. Instead, an easy-to-use stochastic optimization scheme is adopted in this section.

The SA is a metaheuristic approach for global optimization based on the physical analogy with annealing in metallurgy, which aims at reducing the defects of crystals of a material by heating and then cooling it in a controlled way \cite{PJvLaarhoven}. Different from those well-known exact algorithms (e.g., the plain gradient descent method), the SA as a stochastic technique is accepted to be more suitable for addressing hard computational optimization tasks, in a sense that the deterministic schemes can easily fail in such cases.

The procedure of SA is briefly summarized as follows \cite{PJvLaarhoven}. A trial solution point is randomly generated at each iteration, with an acceptance probability, from the current solution point. With the probability of making the transition being specified by a positive acceptance probability function depending on the time-varying temperature parameter and energies of both states, SA not only accepts candidate new points decreasing the objective function (energy) but also allows those increasing it, with some given probability. This is the key to ruling out the local minima where the deterministic algorithms sometimes get stuck. Normally, the SA approach starts with assigning a high value to the temperature, which is then decreased following some annealing schedule as the algorithm proceeds. The extent of search will be reduced as the temperature falls, until a state with the minimum possible energy is reached.

We focus on a more efficient and less hyperparameter dependent variant of the basic SA scheme, known as the adaptive SA (ASA) \cite{OJean,LIngber}, that adaptively and automatically adjusts the temperature and step size at each iteration. Specifically, there are several redesigned functions playing important roles in the implementation of the ASA. The first one is the temperature function $T$ following the annealing schedule: $T_{(k)} = T_{0} \exp \left( - c k^{(1/D)} \right)$, where the subscript $(\cdot)_{(k)}$ indicates the iteration index (starting from scratch), $T_{0}$ is the initially selected temperature, $c$ is a constant controlling the cooling rate, and $D$ denotes the dimension of the parameter space. The second is the generation function $G$ of the parameter vector $\bm{x}$ and temperature $T$, defined as $\tilde{\bm{x}}_{(k)} = G\left(\bm{x}_{(k)}, T_{(k)} \right) = \bm{x}_{(k)} + \bm{s} \circ \bm{r}$,
where $\tilde{\bm{x}}_{(k)}$ denotes the trial solution point generated at the $k$th iteration, $\circ$ represents the Hadamard product, and $\bm{s}$ and $\bm{r}$ are vectors of the same length as $\bm{x}$, containing the scale of the interval which the elements of $\bm{x}$ fall into and random variables in $[-1,1]$ following certain distributions, respectively. The cumulative probability distribution function of the $i$th element of $\bm{r}$ (denoted by $[\bm{r}]_i$) is given as
\begin{equation}{\label{CDF}}
	F_{T_{(k)}}([\bm{r}]_i) = \tfrac{1}{2} +\tfrac{\textup{sgn}\left([\bm{r}]_i\right)}{2} \tfrac{\ln \left( 1 + \left| [\bm{r}]_i \right| / T_{(k)} \right)}{\ln \left( 1 + 1 / T_{(k)} \right)},
\end{equation}
where $\textup{sgn}(\cdot)$ is the signum function. For the purpose of generating values of $[\bm{r}]_i$ according to (\ref{CDF}), the well-known inverse transform sampling method can be employed. It should be pointed out that (\ref{CDF}) implies the dispersion of the random variable $[\bm{r}]_i$ will become smaller as the cooling process goes on \cite{OJean}. Finally, we have the acceptance function:
\begin{align}{\label{Acp}}
&A \left( \tilde{\psi}_{(k)}, \psi_{(k)}, T_{(k)} \right) \nonumber\\
&~= \begin{cases}
	1,~&\tilde{\psi}_{(k)}- \psi_{(k)} \leq 0,\\
	1 \Big/ \left( 1 + \exp{\left( \tfrac{\tilde{\psi}_{(k)} - \psi_{(k)}}{T_{(k)}} \right)} \right),~&\tilde{\psi}_{(k)} - \psi_{(k)} > 0,
	\end{cases}
\end{align}
where $\tilde{\psi}_{(k)}$ is the objective function value associated with the trial solution point generated at the $k$th iteration.

Applying the modified ASA procedure to (\ref{l1_robust}) finally deduces our stochastic framework in Algorithm 2, where $\texttt{rand}$ returns a single uniformly distributed random number in the interval $(0,1)$ and the termination conditions are specified by the maximum number of iterations $N_{\max}$ and another threshold $\gamma > 0$ on the objective function value.

\begin{algorithm}[t]
	\SetAlgoNoLine
	\caption{SA-Based $\ell_1$-Minimization for NLOS Error Mitigation in 3-D AOA Localization.}
	\KwIn{AOA measurements $\{ \hat{\theta}_i \}$ and $\{ \hat{\phi}_i \}$, sensor positions $\{\bm{x}_i\}$, and predefined $T_0$, $N_{\max}$, $\gamma$.}
	
	{
		
		$~~$\textbf{Initialize:} $k \leftarrow 0$, $\bm{x}_{(0)}$ feasibly and randomly, $T_{(0)}$
		$~~$and ${\psi}_{(0)}$ according to the annealing schedule and
		$~~$definition of objective function, respectively, and $~~{\psi}^{\textup{best}} = {\psi}_{(0)}$.

		$~~$\textbf{while}~$k \leq N_{\max}$ and ${\psi}_{(k)} > \gamma$~\textbf{do}

			$~~~~$$\tilde{\bm{x}}_{(k)} = G\left(\bm{x}_{(k)}, T_{(k)} \right)$; 
			
			$~~~~$\textbf{if}$~A \left( \tilde{\psi}_{(k)}, \psi_{(k)}, T_{(k)} \right) \geq \texttt{rand}$

			$~~~~~~$$\bm{x}^{\textup{itm}} \leftarrow \tilde{\bm{x}}_{(k)};~{\psi}^{\textup{itm}} \leftarrow \tilde{\psi}_{(k)};~k \leftarrow k + 1$;
			
			$~~~~~~$\textbf{else~continue}

			$~~~~$\textbf{end}

			$~~~~$\textbf{if}$~{\psi}^{\textup{itm}} < \psi^{\textup{best}}$

			$~~~~~~$$\tilde{\bm{x}} \leftarrow  \bm{x}^{\textup{itm}};~\psi^{\textup{best}} \leftarrow {\psi}^{\textup{itm}}$;

			$~~~~$\textbf{end}
		
	    $~~$\textbf{end}
		
	}
	\KwOut{Estimate of source location $\tilde{\bm{x}}$.}
\end{algorithm}

\section{Numerical Results}
\label{NR}
Numerical results are included in this section to evaluate the performance of Algorithms 1 and 2, in comparison with that of the sole linear LS (termed LLS) and weighted linear LS (termed WLLS) AOA positioning approaches \cite{STomic_LE,STomic6} and an AOA extension from the classical residual weighting algorithm (Rwgh) originally derived for range-based source localization \cite{PCChen}. The basic principle of AOA-based Rwgh here is to introduce additional weights upon (\ref{Residual_LS}) to more prudently combine the intermediate WLS estimates in separate tests, which is summarized in Algorithm 3 for a clearer view. Note that the difference between it and our proposed Algorithm 1 is twofold. First, in Algorithm 3, combinatorial testing is carried out over not just $i$ indicating the index of an observation subset of certain cardinality (just like Algorithm 1), but the number of the used DVOAs $j$. Second, only $N$ DVOAs that minimize the loss in (\ref{Residual_LS}) will be applied to work out the final solution in Algorithm 1, whereas in Algorithm 3, all the available DVOA measurements contribute to the weighted estimation of source position.

\begin{algorithm}[t]
	\SetAlgoNoLine
	\caption{Residual Weighting Data Selection for NLOS Error Mitigation in 3-D AOA Localization (\textbf{for comparison only}).}
	\KwIn{Available DVOAs $\{ \hat{\bm{u}}_i \}$ and sensor positions $\{\bm{x}_i\}$.}
	
	{
		
		$~~$\textbf{Initialize:} $\bm{z} = \bm{0}_3 \in \mathbb{R}^3$ and $\rho = 0$.
		
		$~~$\textbf{for} $j=2,...,L$ (viz., $\mathbb{S}_{2},...,\mathbb{S}_{L}$) \textbf{do}
		
			$~~~~$\textbf{for} $i=1,...,\frac{L!}{j! (L-j)!}$
            
            $~~~~~~$Pass $j$ DVOAs in the $(i,j)$th test to the weighted
            
            $~~~~~~$linear LS method to produce a position estimate
            
            $~~~~~~$$\tilde{\bm{x}}_{j}^{\{i\}}$. Note that the tailoring of $\bm{A}$, $\bm{b}$, and $\bm{W}$ in 
            $~~~~~~$(\ref{WLLS_cfsolu}) follows a similar manner to Algorithm 1.

            $~~~~~~$Based on an indicator for the quality of location
            
            $~~~~~~$estimate, $R\left( j, \tilde{x}_{j}^{\{i\}} \right)$, $\bm{z}$ and $\rho$ are further updated
			
			$~~~~~~$as $\bm{z} \leftarrow \bm{z} + \tfrac{\tilde{\bm{x}}_{j}^{\{i\}}}{{R}\left( j, \tilde{x}_{j}^{\{i\}} \right)}$ and $\rho \leftarrow \rho + \tfrac{1}{{R}\left( j, \tilde{x}_{j}^{\{i\}} \right)}$.

			$~~~~$\textbf{end~for}

	    $~~$\textbf{end~for} with $\tilde{\bm{x}} = \tfrac{\bm{z}}{\rho}$.
		
	}
	\KwOut{Final estimate of source location $\tilde{\bm{x}}$.}
\end{algorithm}

Predefined parameters in Algorithm 2, namely, $T_0$, $N_{\max}$, and $\gamma$, are set to 100, 3000, and $1 \times 10^{-6}$, respectively. The metric for positioning accuracy in the evaluations is the root mean square error (RMSE), defined as $\textup{RMSE} = \sqrt{\frac{1}{N_{\textup{MC}}} \sum_{i=1}^{N_{\textup{MC}}} \left\| \tilde{\bm{x}}^{\{i\}} - \bm{x}^{\{i\}} \right\|_2^2}$, where $N_{\textup{MC}}$ denotes the total of the Monte Carlo (MC) runs, and $\tilde{\bm{x}}^{\{i\}}$ is the estimate of the true source location $\bm{x}^{\{i\}}$ in the $i$th MC run. All the computer simulations and processing of real experimental data are conducted using a laptop with a 4.7 GHz CPU and 16 GB memory.
\subsection{Results of Computer Simulations}
Our 3-D AOA-based source localization scenario comprises a single source to be located and ten sensors with known positions, which are all confined to an origin-centered 20 m $\times$ 20 m $\times$ 20 m cubic room with the locations being randomly chosen in each of $N_{\textup{MC}} = 10000$ ensemble MC trials. The AOA measurements and the corresponding DVOAs are generated according to (\ref{NoisyAOA}) and (\ref{DVOA_def}), respectively. For simplicity, the Gaussian processes in (\ref{CF_general}) are assumed to be of constant standard deviation $\sigma = 1^{\circ}$ for all choices of $i$, whereas $p_i$ equals either $p>0$ or 0, depending on whether the $i$th source-sensor connection is NLOS or LOS. The user-defined parameter for Algorithm 1, i.e., $N$, is fixed at the exact value of the number of LOS links unless otherwise indicated.

\begin{figure}[!t]
	\centering
	\includegraphics[width=3.5in]{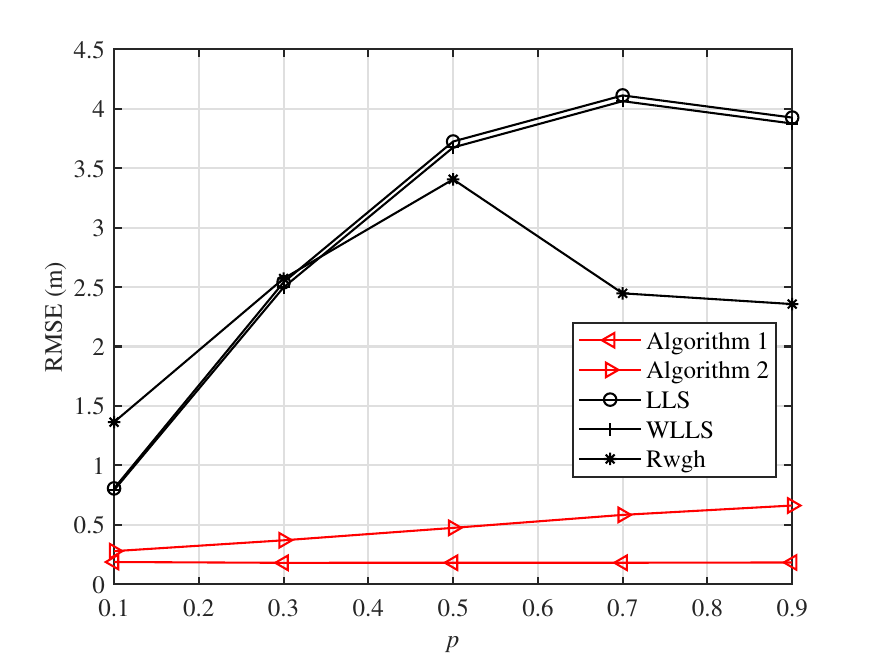}
	\caption{RMSE versus $p$ in mild NLOS scenario.} 
	\label{simu_fig1}
\end{figure}

\begin{figure}[!t]
	\centering
	\includegraphics[width=3.5in]{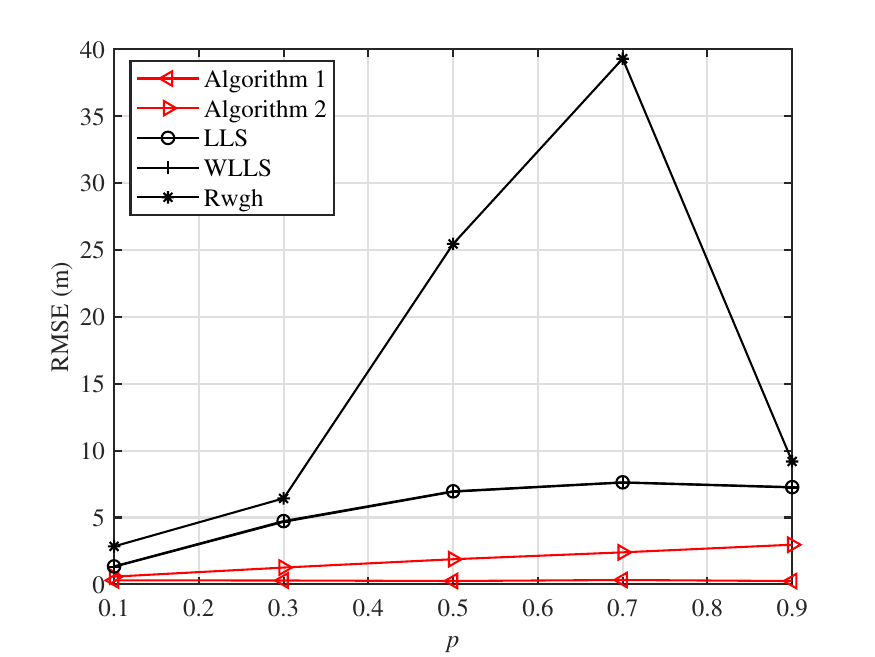}
	\caption{RMSE versus $p$ in moderate NLOS scenario.} 
	\label{simu_fig2}
\end{figure}

\begin{figure}[!t]
	\centering
	\includegraphics[width=3.5in]{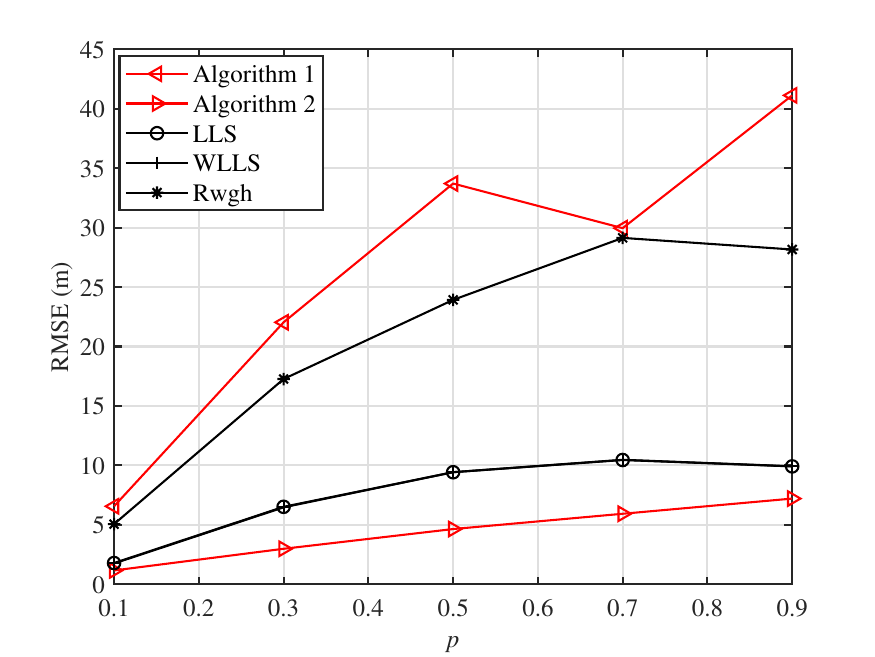}
	\caption{RMSE versus $p$ in severe NLOS scenario.} 
	\label{simu_fig3}
\end{figure}

In our simulations, we consider three typical mixed LOS and NLOS environments, in which two, five, and eight out of ten source-sensor links are subject to $p > 0$, corresponding to the mild, moderate, and severe NLOS propagation conditions, respectively. Figs. \ref{simu_fig1}, \ref{simu_fig2}, and \ref{simu_fig3} demonstrate these comparison results by plotting the RMSE versus $p \in [0.1,0.9]$. Algorithm 1 performs the best and yields very low RMSE values across the whole range of $p$ in the mild and moderate NLOS environments, from which it is deemed that the error-prone connections are successfully identified and discarded in such cases. Because the $\ell_1$ loss only linearly decreases the influence of NLOS errors in the angle measurements, the performance of Algorithm 2 gradually deteriorates over $p$. Nonetheless, it provides a second-best solution after Algorithm 1 and can greatly outperform LLS, WLLS, and Rwgh in these two scenarios. On the other hand, LLS, WLLS, and Rwgh all fail to locate the source in a reliable fashion. Interestingly, the reverse applies in Fig. \ref{simu_fig3} under the severe NLOS conditions. Algorithm 1 performs inferior to its competitors in such a case, since only two links (presumed LOS) are utilized for localization and the estimator in general suffers from the lack of useful location-bearing information. In contrast, Algorithm 2 leveraging the $\ell_1$-minimization criterion is still robust and capable of producing the lowest RMSE results this time.

There are also other notable results from the comparison of Figs. \ref{simu_fig1}--\ref{simu_fig3}. For example, the performance gap between LLS and WLLS disappears as the NLOS propagation conditions become severer, namely the number of erroneous paths increases. Relying on the use of (\ref{Residual_LS}) similar to our Algorithm 1, the Rwgh scheme (viz., Algorithm 3) however might not reliably mitigate the adverse effects of NLOS errors in the AOA context, as it does in distance-based positioning \cite{PCChen,WXiong5}. This is particularly so in the moderate NLOS scenario shown in Fig. \ref{simu_fig2}, where a more drastic fluctuation in the RMSE delivered by Rwgh is observed compared to Figs. \ref{simu_fig1} and \ref{simu_fig3}. The results imply that introducing weights to combinations based on the normalized residuals can still lead to largely biased AOA location estimates in many cases and, therefore, might not be a fine option. A possible improvement strategy is to replace the normalized residual in Rwgh by the probability obtained from statistical hypothesis tests \cite{SWu}. We only provide here an outlook, since the detailed analysis of an improved Rwgh approach is far beyond the scope of this study.

\begin{figure}[!t]
	\centering
	\includegraphics[width=3.5in]{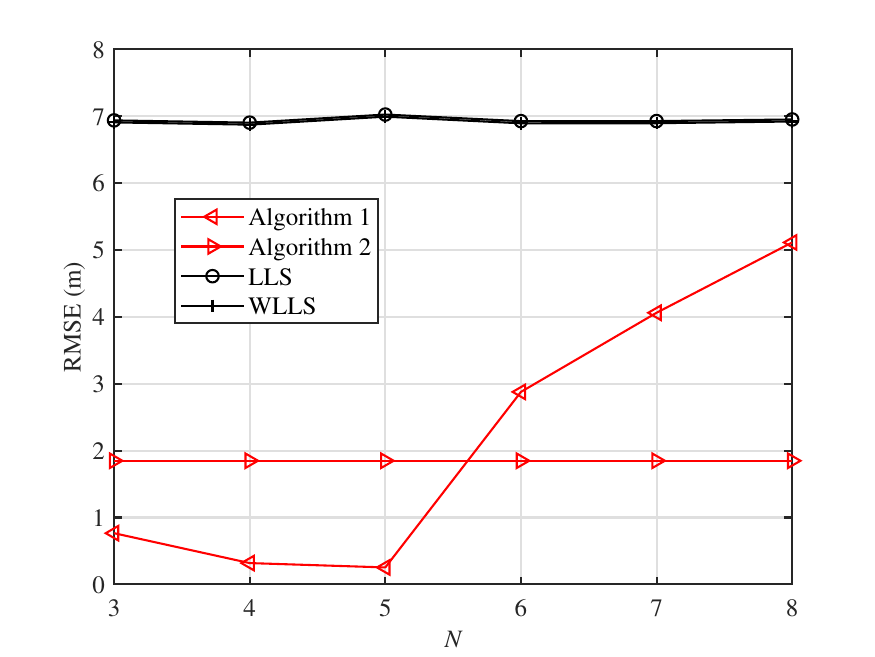}
	\caption{RMSE versus $N$ in moderate NLOS scenario.} 
	\label{simu_fig4}
\end{figure}

Additionally, we investigate the impact of $N$ on the localization performance of Algorithm 1. Fig. \ref{simu_fig4} plots the RMSE versus $N$ for Algorithms 1 and 2, LLS, and WLLS in the moderate NLOS scenario above, where the results of Rwgh are excluded as they in general do not show much significance. It is seen that the performance degradation when $N$ is underestimated ($N < 5$) is much milder than the cases when it is overestimated ($N > 5$), as long as the AOA measurements used for positioning remain adequate. In overestimated situations, the performance of Algorithm 1 will gradually deteriorate towards that of LLS and WLLS, as $N$ increases.

To summarize, it is preferred employing Algorithm 1 if the number of NLOS connections is not large enough and at least roughly known, whereas one may simply resort to the $\ell_1$-loss-based robust scheme in cases with only limited \textit{a priori} NLOS information, just to be on the safe side.

\subsection{Results of Real-World Experiments}
\begin{figure}
	\centering
	\includegraphics[width=\columnwidth]{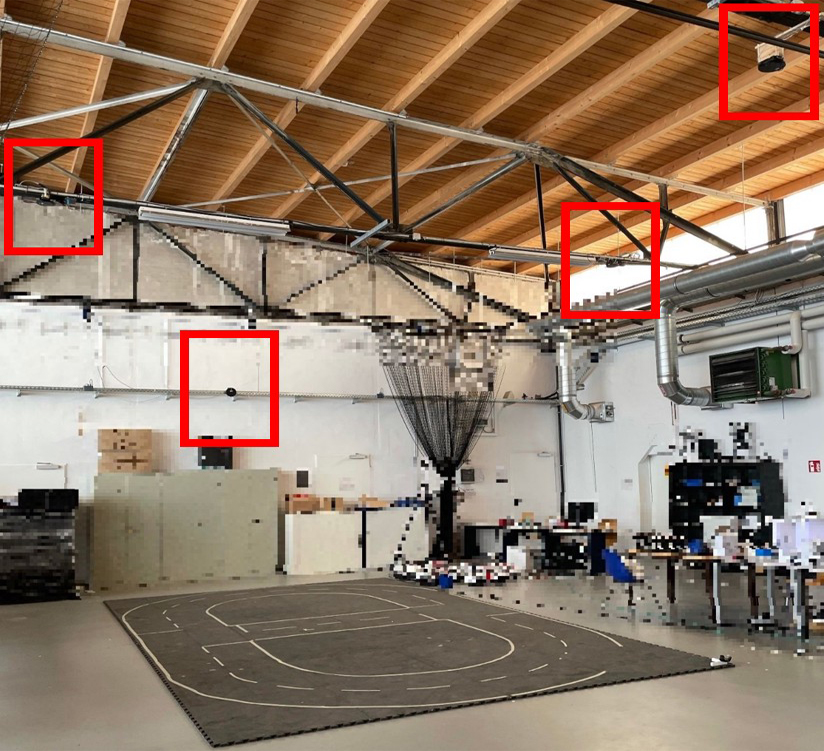}
	\caption{Experimental setup. AOA receivers are marked with red rectangles.}
	\label{fig:hangar_photo}
\end{figure}

As shown in Fig. \ref{fig:hangar_photo}\footnote{Note that pixelization was applied in Fig. \ref{fig:hangar_photo} to cover the equipment from other laboratories for confidentiality purposes.}, ultrasonic onsite experiments were carried out in the hangar at the Technische Fakult\"{a}t campus of the University of Freiburg, Freiburg, Germany, where a Soberton Inc. SP-1303L speaker was placed in several different positions on the ground beneath the AOA receivers. Each AOA receiver is equipped with five TDK-InvenSense ICS-40720 microphones. Both speaker and microphones are commercial off-the-shelf hardware, which have been tested in laboratory to suit our needs. For the experiments, we employed frequency modulated chirp signals of duration $T_\textrm{s} = 50~\textrm{ms}$, start frequency $f_\textrm{start} = 20~\textrm{kHz}$, end frequency $f_\textrm{end} = 40~\textrm{kHz}$, and exponential instantaneous frequency. A chirp pulse train was played for two minutes with a pause in between two consecutive chirp pulses of $T_\textrm{pause} = 250~\textrm{ms}$.
For each pulse, the DVOA was calculated for each receiver and fed into our proposed algorithms to estimate the source position. The orientation of the AOA receivers was calibrated by fusing the time-difference-of-arrival and AOA measurements. During the calibration process, the source moved into the localization area with a fixed height. Constrained optimization was then used to estimate the directions to which the receivers were pointing.

In such a setting, four AOA receivers were placed at known positions with known orientation. A total station theodolite was used to estimate their true positions, which are given in Table \ref{tab:aoa_receiver_coordinates}, as well as five reference points beneath the AOA receivers for evaluation purposes.

\begin{table}[ht]
    \setcellgapes{5pt}
    \centering
    \caption{AOA Receiver (Rec\#) and Reference Point (RP\#) Coordinates}
    \label{tab:aoa_receiver_coordinates}
    \begin{tabular}{lccc|lccc}
        \toprule
        \thead{Index} &\thead{x (m)}&\thead{y (m)}&\thead{z (m)}&\thead{Index} &\thead{x (m)}&\thead{y (m)}&\thead{z (m)} \\
        \midrule
        Rec1 & 1.99 & -1.20 & 4.65 & RP1 & 3.03 & 7.33 & 0 \\
        Rec2 & -2.15 & 3.69 & 4.74 & RP2 & 0.83 & 7.27 & 0 \\
        Rec3 & 4.75 & 3.73 & 4.78 & RP3 & -1.33 & 7.26 & 0 \\
        Rec4 & 0.87 & 8.57 & 3.35 & RP4 & -2.09 & 3.69 & 0 \\
        -&-&-&-&RP5 & -2.09	& 2.36 & 0\\
        \midrule
        \bottomrule
    \end{tabular}
\end{table}

\begin{figure}[!t]
	\centering
	\includegraphics[width=3.5in]{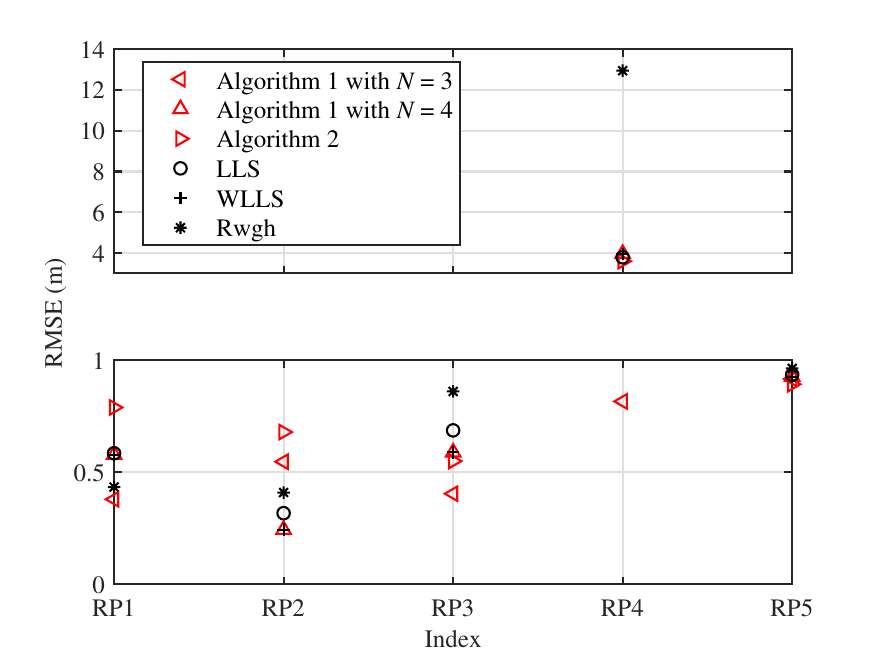}
	\caption{RMSE versus reference point index.} 
	\label{exp_fig1}
\end{figure}

\begin{figure*}[!t]
	\centering
	\includegraphics[width=7in]{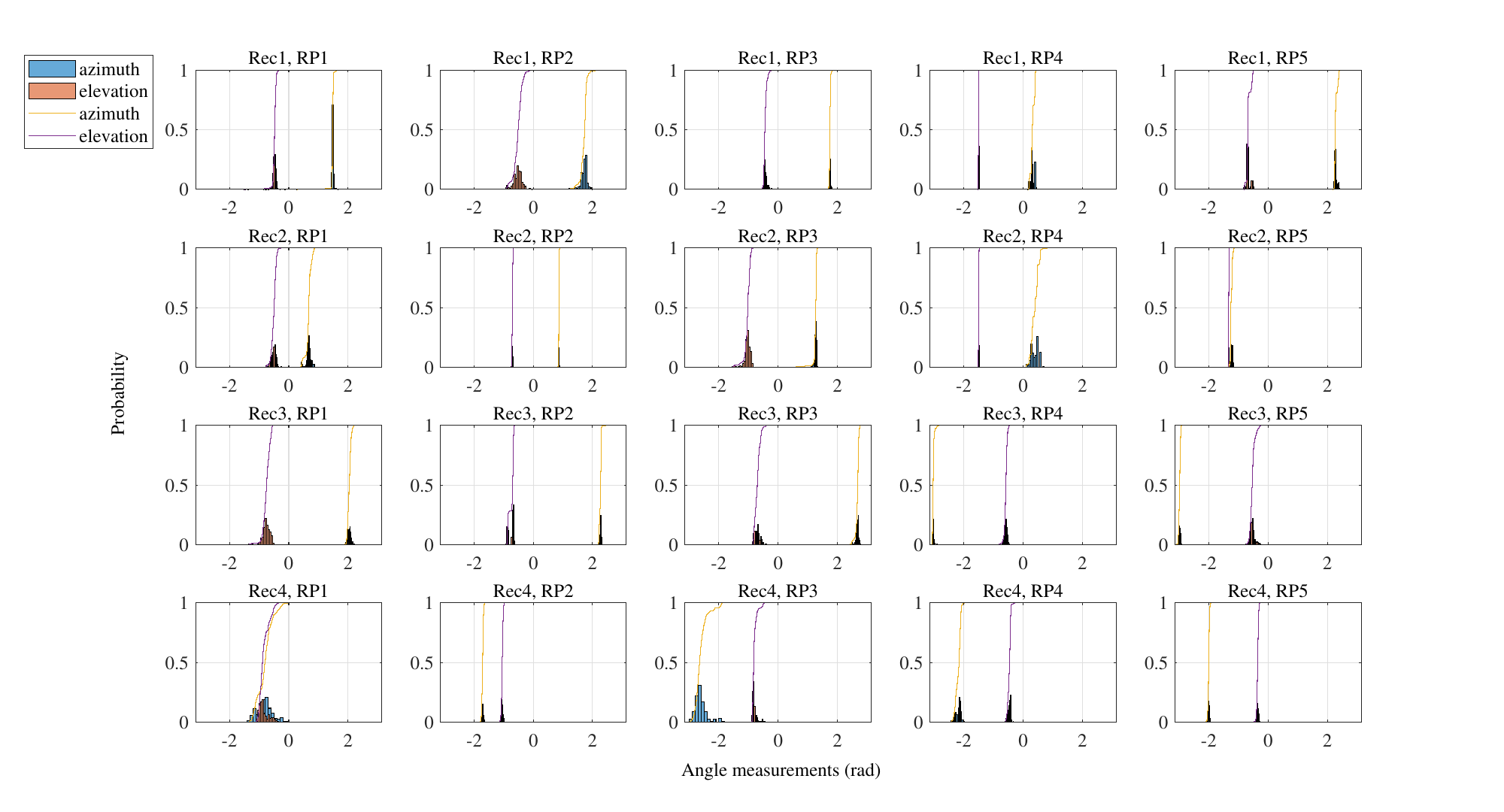}
	\caption{Histogram (using probability normalization) and empirical CDF plots of azimuth and elevation angle measurements based on 500 samples.} 
	\label{exp_fig2}
\end{figure*}

Rec1-Rec3 were mounted to the ceiling at nearly the same height pointing towards the ground, and Rec 4 was placed on a vertical wall pointing to the center of the hall. Under multiple adverse environmental conditions including but not limited to obstacles and reflecting surfaces, disturbance and errors from acoustic echoes might occur in the sensor-collected signal observations. Fig. \ref{exp_fig1} illustrates the RMSE for different algorithms corresponding to the histogram and empirical cumulative distribution function (CDF) plots of angle measurements in Fig. \ref{exp_fig2}, based on $500$ MC samples acquired at each RP. It is observed from Fig. \ref{exp_fig1} that Algorithm 1 with $N=3$ yields the best results for RP1, RP3, and RP4, which agrees with the numerical data distributions showcased in Fig. \ref{exp_fig2}, since there is always one path subject to comparatively severer noise/error conditions (i.e., (Rec4, RP1), (Rec4, RP3), and (Rec2, RP4), respectively) than the other three in each of the corresponding localization scenarios. For speaker placed at RP5 where a few outliers exist in the angle measurements collected by Rec1, all the considered methods have comparable performance and Algorithm 2 produces slightly better accuracy than the others. The case of RP2 is a bit more complicated, as it can be seen from Fig. \ref{exp_fig2} that not only does the (Rec3, RP2) path data contain outliers but also the (Rec1, RP2) counterparts are immersed in higher-level noise/errors. In such circumstances, Algorithm 1 with $N=3$, Algorithm 2, and Rwgh all show no improvements over the plain LS estimators, whereas Algorithm 1 with $N=4$ coincides with WLLS and delivers the minimum RMSE.

\section{Conclusion}
\label{CC}
In this paper, we have put forward two efficient and easy-to-use error mitigation algorithms for 3-D AOA source localization in the mixed LOS/NLOS indoor environments. The first method is built upon a weighted linear LS solution and the procedure of residual-based data selection, whereas the second implements a statistically robust location estimator by stochastically dealing with the $\ell_1$-minimization formulation. Using the synthetic and real experimental data, we have demonstrated the superior efficacy of two proposed approaches in diverse representative mixed LOS/NLOS environments. We have also illustrated that the presented data-selective LS and SA-assisted $\ell_1$-minimization schemes can benefit from the completeness of removal of erroneous links in mild/moderate NLOS scenarios and the overall robustness, respectively. Moreover, a distinct advantage of our two methods is that neither of them require precise prior information about NLOS errors. A possible future direction can be the development of other computationally more appealing solvers for $\ell_1$-minimization in this context.

\end{document}